\documentclass[english,aps,preprint]{revtex4}
\usepackage[T1]{fontenc}
\usepackage[latin9]{inputenc}
\pdfoutput=1
\usepackage{amsmath}
\usepackage{amssymb}
\usepackage{graphicx}

\makeatletter

\newcommand{\noun}[1]{\textsc{#1}}

\@ifundefined{textcolor}{}
{%
 \definecolor{BLACK}{gray}{0}
 \definecolor{WHITE}{gray}{1}
 \definecolor{RED}{rgb}{1,0,0}
 \definecolor{GREEN}{rgb}{0,1,0}
 \definecolor{BLUE}{rgb}{0,0,1}
 \definecolor{CYAN}{cmyk}{1,0,0,0}
 \definecolor{MAGENTA}{cmyk}{0,1,0,0}
 \definecolor{YELLOW}{cmyk}{0,0,1,0}
}

\usepackage{babel}

\makeatother

\usepackage{babel}
\begin{document}

\preprint{This line only printed with preprint option}

\title{Energy Flux in Hierarchical Equations of Motion Method and Its Application
to a Three-Level Heat Engine}

\author{Yun ZHOU}
\email{zhouyun@gznc.edu.cn}

\affiliation{Guizhou Provincial Key Laboratory of Computational Nano-material
Science, Guizhou Synergetic Innovation Center of Scientific Big Data
for Advanced Manufacturing Technology, Guizhou Education University,
Guiyang, Guizhou, 550018, China }

\author{Jun CAO}
\email{caojunbnu@mail.bnu.edu.cn}

\affiliation{School of Materials Science and Energy Engineering, Foshan University,
Foshan, Guangdong, 528000, China}

\author{Yun-An YAN}
\email{yunan@ldu.edu.cn}

\affiliation{School of Physics and Optoelectronic Engineering, Ludong University,
Yantai, Shandong, 264025, China }

\author{Tie-Jun XIAO}
\email{tjxiao@gznc.edu.cn}

\affiliation{Guizhou Provincial Key Laboratory of Computational Nano-material
Science, Guizhou Synergetic Innovation Center of Scientific Big Data
for Advanced Manufacturing Technology, Guizhou Education University,
Guiyang, Guizhou, 550018, China }

\maketitle
An accurate numerical method that calculates energy flux of quantum
open system is useful for the investigation of heat transport and
work generation in small quantum systems. We derive the formula of
energy flux in the framework of hierarchical equations of motion (HEOM)
method with the help of stochastic decoupling technique. The resulting
expression is a combination of the terms in the second layer of the
hierarchy. The formula is applied to a three-level heat engine coupled
to three baths, of which two for heat sources and one for work dump.
We illustrate the proper parameterizing to converge the third bath
to the ``work-dump'' limit. As an example, the effect of the engine
parameters on the working efficiency is studied.

\section{introduction}

Due to the development of technologies there have been observed in
many biological systems that microscopic devices provide similar utilities
as normal-size engines do. For instance the photosynthetic reaction
centers of plants and bacteria produce free energy from two heat sources,
the sun and the cold ambiance of the earth, that resembles the functionality
of a heat engine\cite{gelbwaser-klimovsky2017onthermodynamic}. For
microscopic machines fueled by free energy, it is also important to
develop a perspective of thermodynamics. Despite the obvious pragmatic
purposes of extracting useful utilities, theory of heat engines shares
the common ground of entropy/information with the active research
fields of quantum information transmission and computation. Work extraction
of heat engines is a major connection bridging the mathematical concept
of information and the concrete reality. Studying quantum models of
heat engines may give another angle of understanding these fundamental
concepts.

Microscopic heat engines are different from their normal-size counterparts.
The investigation of heat engines laid its ground on the pursuit of
work. In normal scale, work is unambiguously defined as the product
of force and displacement, the two fundamental concepts of Newtonian
mechanics, and can be transformed by ideal machineries to various
forms without changing its quantity. In microscopic scale, however,
the fluctuation in force and displacement caused by classical randomness
and quantum uncertainty makes the deterministic definition inappropriate.
More substantially, the interaction of microscopic heat engines with
their environments could be much stronger than their normal-size counterparts.
Unlike the classical theory in which the engine is well separated
from the environment and interact weakly to the almost unperturbed
heat baths, the microscopic heat engine is virtually embeded into
the heat baths. The interaction heavily disturbs both the engine and
the heat baths that per se breaks the assumed settings of the classical
theory. It is therefore neccessary to bring in more fundamental models
and dynamic viewpoints into the study of microscopic heat engines.

There have been various microscopic models of heat engines. Some,
like Szilard's engine \cite{szilard1926onthe} and Feynman's ratchet
and pawl \cite{richardfeynman1963thefeynman}, are heuristic in giving
explicit information mechanisms. Close examinations upon them demonstrates
the necessity of integrating information processes (e.g. measurement)
into their thermodynamics \cite{brillouin2004science}. Also inferred
is that information is as well restricted by the laws of thermodynamics
\cite{landauer1961irreversibility}. Despite the conceptual importance,
their dynamics is not easy to simulate. Some recent efforts have made
them more accessible to strict analysis \cite{magnasco1998feynmans,hanggi2009artificial,zhou2010minimal,mandal2012workand,quan2007quantum}.
In contrast to these heuristic models, there are also engines much
easier to simulate. A pioneering one is a model of maser proposed
by Scovil and Schulz-Dubios \cite{scovil1959threelevel} where a simple
three-level system is coupled to two heat baths and one work dump.
The model is equivalent to a heat engine and subject to the same Carnot
efficiency. This discovery has ignited many following researches \cite{geva1994threelevel,boukobza2007threelevel,allahverdyan2016adaptive}.
Simple as they are, the informational mechanisms of them are not obvious.
We leave such theoretic inquiries to theorists and in the present
work restrict ourselves to the construction of a reliable numerical
method that calculates work and heat flows for such models.

A standard method dealing with such models is the master equation
pioneered by Lindblad \cite{lindblad1976onthe}, Gorini, Kossakowski
and Sudarshan \cite{gorini1976completely}. It is then applied to
quantum heat engines by Kosloff \cite{kosloff1984aquantum}. The method
consistently reproduces the laws of equilibrium thermodynamics \cite{kosloff2013quantum}
which qualifies it as an eligible generalization of thermodynamics
to quantum regime. Despite its quantum nature, the method depends
on idealizations like weak coupling limit and Markovian approximation.
These assumptions may be inappropriate because microscopic heat engines
could be tightly embeded into its environments and the time scale
separation could be invalid for systems varying as fast as the fluctuations
of their heat baths. Moreover, the necessity of these approximations
is not seen in the elegant informational mechanisms of microscopic
heat engines. Hierarchical equations of motion (HEOM) method is a
powerful tool to go beyond the approximations\cite{tanimura1990nonperturbative,yan2004hierarchical,xu2005exactquantum,shi2009efficient}.
HEOM is a numerically exact and remarkably efficient tool to simulate
models of small quantum systems coupled to harmonic heat baths. Though
its bath model is limited by the requirement of harmonicity, its validity
has been argued \cite{caldeira1983quantum} and tested in many researches.
One way of deriving HEOM is through stochastic decoupling \cite{yan2004hierarchical},
which is also the route the present research takes. Stochastic decoupling
provides rigorous mathematics to separate two subsystems connected
by the interaction of factorized form. The separation allows to calculate
the local observables of the interested subsystem. With the above-mentioned
tools we find the expression of energy flux in HEOM formalism and
then apply it to a three-level heat engine. The approach does not
calculate full counting statistics as does in Ref \cite{cerrillo2016nonequilibrium},
therefore assumes a simpler form.

The rest of the paper is arranged as follows. Section \ref{sec:The-expression-of}
derives the expression of energy flux for the system-plus-bath models
in the framework of HEOM. Section \ref{sec:Theory-for-a} applies
the general formalism to a three-level heat engine coupled to three
baths. The proper parameterizing to converge one of the baths to a
work dump is illustrated. Section \ref{sec:The-efficiency-of} exemplifies
the method by calculating the efficiency of the heat engine.

\section{The expression of energy flux in HEOM\label{sec:The-expression-of}}

Consider a system-plus-bath complex $H=H_{s}+H_{b}+f_{s}g_{b}$, where
$H_{s}$ and $H_{b}$ stand for the system Hamiltonian and the bath
Hamiltonian respectively. The interaction between the system and the
bath is a simple product of a system operator $f_{s}$ and a bath
operator $g_{b}$. The basis of defining a flux is that a flux into
the system increases its respective physical quantity. Calling the
operator of the quantity $A_{s}$, we define its flux $\left\langle j\right\rangle =\frac{d}{dt}\left\langle A_{s}\right\rangle $.
Using the propagator of the Hamiltonian, it is straightforward to
show that $\left\langle j\right\rangle =Tr\left\{ \frac{i}{\hbar}\left[H_{s},A_{s}\right]\rho\left(t\right)\right\} +Tr\left\{ \frac{i}{\hbar}g_{b}\left[f_{s},A_{s}\right]\rho\left(t\right)\right\} $
, where the flux operator is identified as $j=\frac{i}{\hbar}\left[H_{s},A_{s}\right]+\frac{i}{\hbar}g_{b}\left[f_{s},A_{s}\right]$
and $\left[\cdot,\cdot\right]$ is the commutator. The factorization
of system operator and bath operator in the expression naturally sees
the application of stochastic decoupling technique \cite{shao2004decoupling}.
To perform that, first we note the total density matrix can be cast
in the form $\rho\left(t\right)=M_{u_{1},u_{2}}\left\{ \rho_{s}\left(u_{1},u_{2}^{*}\right)\rho_{b}\left(u_{1}^{*},u_{2}\right)\right\} $,
where the operator $M_{u_{1},u_{2}}\left\{ \cdot\right\} $ averages
its operant over the white noises $u_{1}$ and $u_{2}$. Here $\rho_{s}$
and $\rho_{b}$ are operators of, respectively, system space and bath
space, both driven by the noises $u_{1}$ and $u_{2}$: 
\[
i\hbar d\rho_{s}\left(t\right)=\left[H_{s},\rho_{s}\left(t\right)\right]dt+\frac{1}{2}\left[f_{s},\rho_{s}\left(t\right)\right]u_{1,t}dt+\frac{i}{2}\left\{ f_{s},\rho_{s}\left(t\right)\right\} u_{2,t}^{*}dt
\]
\[
i\hbar d\rho_{b}\left(t\right)=\left[H_{b},\rho_{b}\left(t\right)\right]dt+\frac{1}{2}\left[g_{b},\rho_{b}\left(t\right)\right]u_{2,t}dt+\frac{i}{2}\left\{ g_{b},\rho_{b}\left(t\right)\right\} u_{1,t}^{*}dt.
\]
It is easy to validate with Itô calculus that $i\hbar\frac{d}{dt}\rho=[H,\rho]$.
With the factorized form the flux is written as 
\begin{equation}
\left\langle j\right\rangle =\frac{i}{\hbar}Tr_{s}\left\{ \left[H_{s},A_{s}\right]\tilde{\rho}_{s,0}\right\} +\frac{i}{\hbar}Tr_{s}\left\{ \left[f_{s},A_{s}\right]\tilde{\rho}_{s,1}\right\} ,\label{eq:flux_expr}
\end{equation}
where the bath plays its role via 
\begin{align}
\tilde{\rho}_{s,1} & =M_{u_{1},u_{2}}\left\{ \rho_{s}Tr_{b}\left\{ \rho_{b}\right\} \bar{g}\left(t\right)\right\} \label{eq:density_matrix_of_flux}\\
\tilde{\rho}_{s,0} & =M_{u_{1},u_{2}}\left\{ \rho_{s}Tr_{b}\left\{ \rho_{b}\right\} \right\} .\nonumber 
\end{align}
Notice $\tilde{\rho}_{s,0}$ is actually the reduced density matrix
$\tilde{\rho}_{s}$ of the system and $\tilde{\rho}_{s,1}$ is some
correlation between the system and the random force $\bar{g}\left(t\right)=Tr_{b}\left\{ \rho_{b}g_{b}\right\} /Tr_{b}\left\{ \rho_{b}\right\} $
exerted by the bath. For the bath of Caldeira-Leggett model $g_{b}=\sum_{j}c_{j}x_{j}$
and $H_{b}=\sum_{j}\left(\frac{1}{2}m_{j}\omega_{j}^{2}x_{j}^{2}+\frac{p_{j}^{2}}{2m_{j}}\right)$
\cite{caldeira1983quantum}, $\rho_{b}$ is solved analytically in
the form $Tr_{b}\left(\rho_{b}\right)=\exp\left(\int_{0}^{t}\bar{g}\left(t^{'}\right)u_{1,t^{'}}^{*}dt^{'}/\sqrt{\hbar}\right)$
where $\bar{g}\left(t\right)=\sqrt{\hbar}\int_{0}^{t}\alpha_{I}\left(t-t^{'}\right)u_{2,t^{'}}dt+\sqrt{\hbar}\int_{0}^{t}\alpha_{R}\left(t-t^{'}\right)u_{1,t^{'}}^{*}$
and $\alpha_{I(R)}\left(t\right)$ is the imaginary(real) part of
the response function $\alpha\left(t\right)=\frac{1}{\pi}\int_{0}^{\infty}d\omega\,J\left(\omega\right)\left[\coth\left(\beta\hbar\omega/2\right)\cos\left(\omega t\right)-i\sin\left(\omega t\right)\right]$
. Here $J\left(\omega\right)=\frac{\pi}{2}\sum_{j}\frac{c_{j}^{2}}{m_{j}\omega_{j}}\delta\left(\omega-\omega_{j}\right)$
is the spectral density function of the bath. Because of its linear
form in the noise $u_{1}^{*}$, $Tr_{b}\left\{ \rho_{b}\right\} $
is absorbed into the weight function of the noises in Eq. (\ref{eq:density_matrix_of_flux})
with Girsanov transformation, which simplifies the expressions to:
\begin{align}
\tilde{\rho}_{s,1} & =M_{u_{1},u_{2}}\left\{ \bar{\rho}_{s}\left(t\right)\bar{g}\left(t\right)\right\} \label{eq:redu_rho}\\
\tilde{\rho}_{s,0} & =M_{u_{1},u_{2}}\left\{ \bar{\rho}_{s}\right\} .\nonumber 
\end{align}
The transform also affects $\rho_{s}$ and results in a new density
matrix $\bar{\rho}_{s}$ driven by the new SDE 
\begin{equation}
i\hbar d\bar{\rho}_{s}=\left[H_{s},\bar{\rho}_{s}\right]dt+\left[f_{s},\bar{\rho}_{s}\right]\bar{g}\left(t\right)dt+\frac{1}{2}\left[f_{s},\bar{\rho}_{s}\right]u_{1}dt+\frac{i}{2}\left\{ f_{s},\bar{\rho}_{s}\right\} u_{2}^{*}dt.\label{eq:sto_rho}
\end{equation}
Averaging the equation gives formally $i\hbar\frac{d}{dt}\tilde{\rho}_{s,0}=\left[H_{s},\tilde{\rho}_{s,0}\right]+\left[f_{s},\tilde{\rho}_{s,1}\right]$,
which straightforwardly validates Eq. (\ref{eq:flux_expr}) . The
key quantities $\tilde{\rho}_{s,0}$ and $\tilde{\rho}_{s,1}$ could
in principle be calculated by propagating and averaging the SDE (Eq.(\ref{eq:sto_rho})).
However, in practice it is indeed obtained via the more efficient
HEOM through the connection between the SDE and HEOM\cite{yan2004hierarchical}.
The connection is sketched out at the end of this section.

For energy flux $A_{s}=H_{s}$, Eq. (\ref{eq:flux_expr}) is reduced
to 
\begin{equation}
\left\langle j\right\rangle =\frac{i}{\hbar}Tr_{s}\left\{ \left[f_{s},H_{s}\right]\tilde{\rho}_{s,1}\right\} .\label{eq:energy_flux_j}
\end{equation}
A slight different way to define energy flux is $\left\langle j^{'}\right\rangle =-\frac{d}{dt}$$\left\langle H_{b}\right\rangle $.
The two ways are nonequivalent as, by energy conservation, $\left\langle j^{'}\right\rangle =\left\langle j\right\rangle +\frac{d}{dt}\left\langle f_{s}g_{b}\right\rangle $.
There is a difference $\frac{d}{dt}\left\langle f_{s}g_{b}\right\rangle $
between the two definitions. The latter definition is also computable
with similar procedures, only a little more complicated. To get $\frac{d}{dt}\left\langle f_{s}g_{b}\right\rangle $
we start from $\left\langle f_{s}g_{b}\right\rangle =Tr\left\{ \rho f_{s}g_{b}\right\} $
and $\rho\left(t\right)=M_{u_{1},u_{2}}\left\{ \rho_{s}\left(u_{1},u_{2}^{*}\right)\rho_{b}\left(u_{1}^{*},u_{2}\right)\right\} $.
Combining the equations above we have $\left\langle f_{s}g_{b}\right\rangle =Tr_{s}\left\{ f_{s}M_{u_{1},u_{2}}\left\{ \rho_{s}Tr\left(\rho_{b}\right)\bar{g}\left(t\right)\right\} \right\} $,
$Tr\left(\rho_{b}\right)$ is as usual absorbed with the help of Girsanov
transformation to get $\left\langle f_{s}g_{b}\right\rangle =Tr_{s}\left\{ f_{s}\tilde{\rho}_{s,1}\right\} $.
Adding the two parts, the expression for the second definition of
energy flux is revealed to be 
\begin{equation}
\left\langle j^{'}\right\rangle =\frac{i}{\hbar}Tr_{s}\left\{ \left[f_{s},H_{s}\right]\tilde{\rho}_{s,1}\right\} +Tr_{s}\left\{ f_{s}\frac{d}{dt}\tilde{\rho}_{s,1}\right\} .\label{eq:flux_expr1}
\end{equation}

From the SDE (Eq. (\ref{eq:sto_rho})), HEOM is derived by working
out the ordinary differential equation (\emph{\noun{ODE}}) of the
average $M\left\{ \bar{\rho}_{s}\right\} $ with the help of Itô Calculus
\cite{yan2004hierarchical}. As has been intensively studied and applied
to many systems \cite{yan2012lasercontrol,ishizaki2009theoretical,chen2010twodimensional},
HEOM is effective and precise to simulate quantum dissipative systems
with fast decaying response functions. Simply saying, the technique
splits the response function $\alpha_{I,R}\left(t\right)$ into the
sum of exponential functions. Respectively $\bar{g}\left(t\right)$
is also divided into the sum of $\bar{g}_{j}\left(t\right)$s, each
of which is for one exponential. The collection of averages $M_{u_{1},u_{2}}\left\{ \bar{\rho}_{s}\Pi_{j}\bar{g}_{j}\left(t\right)^{n_{j}}\right\} $,
labeled by indexes $\left(n_{1},n_{2},\cdots\right)$ ($n_{j}$ is
a non-negative integer), is determined by a hierarchically related
set of ODEs. Out of them $M_{u_{1},u_{2}}\left\{ \bar{\rho}_{s}\right\} $
(the reduced density matrix) is the leading term with $\left(n_{1},n_{2},\cdots\right)=\left(0,0,\cdots\right)$.
The hierarchy is truncated to finite layers $\sum_{j}n_{j}<N$ for
numerical practicability. The cross relations of the coupled ODEs
are sparse because any term labeled by $\left(n_{1},n_{2},\cdots\right)$
is related only to its immediate superior and subordinate terms $\left(n_{1},n_{2},\cdots,n_{j}\pm1,n_{j+1},\cdots\right)$.
This feature contributes to its high numerical efficiency. In many
applications, only the leading term $\tilde{\rho}_{s,0}=M_{u_{1},u_{2}}\left\{ \bar{\rho}_{s}\right\} $
of HEOM is required to calculate the system-related quantities. But
for fluxes, $\tilde{\rho}_{s,1}$ (Eq. (\ref{eq:energy_flux_j}))
and probably $\frac{d}{dt}\tilde{\rho}_{s,1}$ (Eq. (\ref{eq:flux_expr1}))
are also needed. It is seen from Eq. (\ref{eq:redu_rho}) that $\tilde{\rho}_{s,1}=\sum_{j}M_{u_{1},u_{2}}\left\{ \bar{\rho}_{s}\left(t\right)\bar{g}_{j}\left(t\right)\right\} $
is the sum of the second layer terms of HEOM (with $\sum_{j}n_{j}=1$
). Its derivative $\frac{d}{dt}\tilde{\rho}_{s,1}$ is also known
to be a certain combination of the first layer ($\sum_{j}n_{j}=0$)
and the third layer ($\sum_{j}n_{j}=2$) terms based on the hierarchical
dependence of the ODEs. The exact form of $\frac{d}{dt}\tilde{\rho}_{s,1}$
relies on the shapes and the specific splitting of $\alpha_{I,R}\left(t\right)$,
thus is not presented generally here. For convenience, we choose the
first definition Eq. (\ref{eq:flux_expr}) in thereafter simulations
because of its independent form and lesser requirements. The dependence
of transportation properties of fermion system on the second layer
terms of HEOM is also found in Ref. \cite{jin2008exactdynamics}.

\section{Application to a three-level heat engine\label{sec:Theory-for-a}}

\begin{figure}
\includegraphics[scale=0.5]{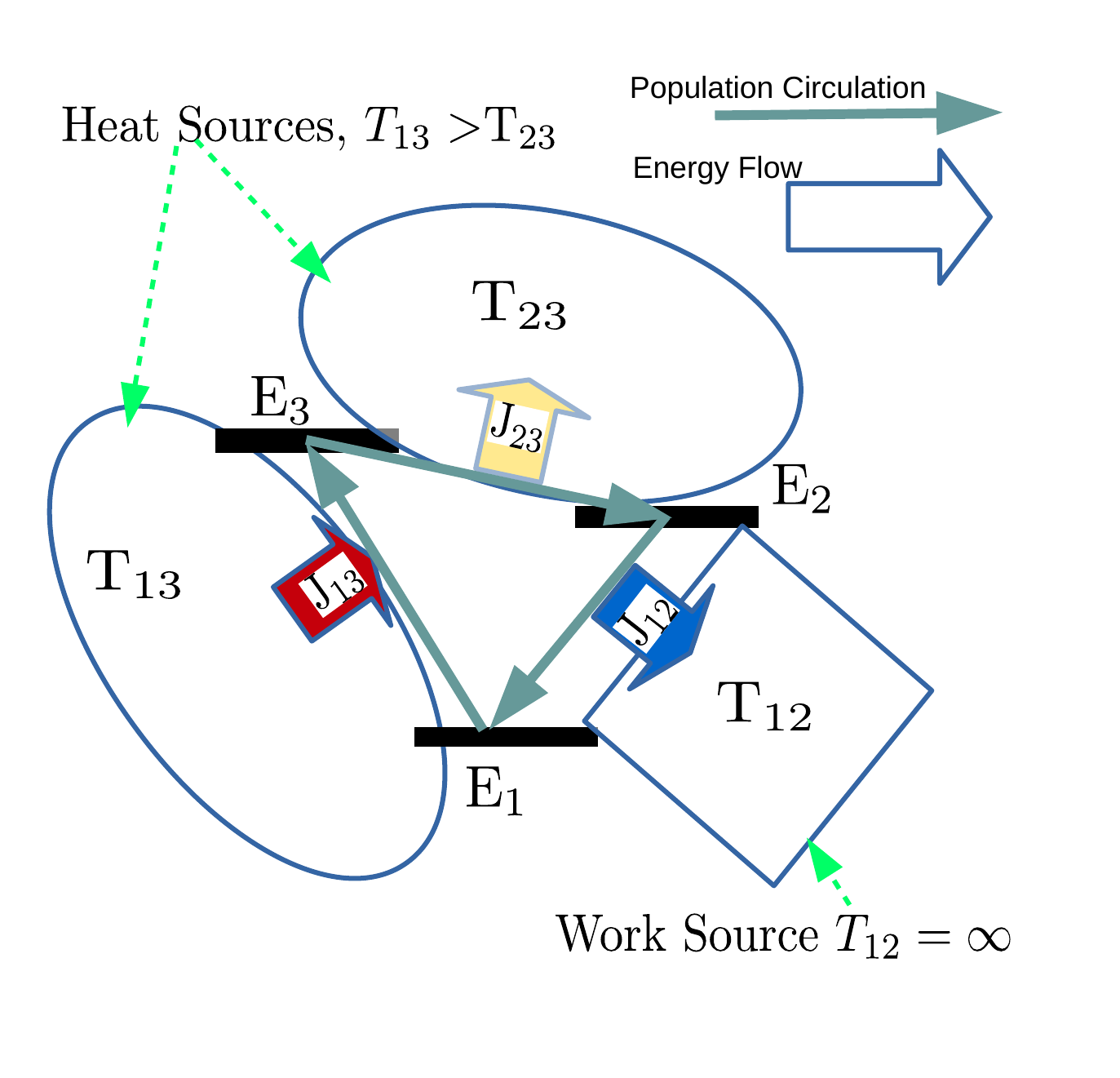}\caption{The schematic plot of the three-level heat engine with three system
states $E_{j}$, three baths labeled by their temperatures $T_{ij}$
and three energy fluxes $J_{ij}$. The square stands for the work
dump and the two ovals for heat baths. \label{fig:The-schematic-plot}}
\end{figure}

The three-level heat engine we study is illustrated in Figure \ref{fig:The-schematic-plot}.
A quantum system with three states (labeled by their energy $E_{j}$)
is coupled to three baths (labeled by their temperature $T_{ij}$).
Each bath causes the transition between two of the states ($E_{i}$
and $E_{j}$) and produces an energy flux $J_{ij}$ in or out of the
bath. We choose the incoming flux to the system to be positive and
the other direction negative. The baths are Caldeira-Leggett models
with friction coefficients $\eta_{ij}$. This model allows for a full
quantum description with time-independent Hamiltonian and thus provides
some rigors in discussing subtle quantum effects.

\subsection{HEOM for a three-level system}

To model the heat engine named above, we need to couple three Caldeira-Leggett
baths to the three-level engine Hamiltonian to transit between its
levels, the Hamiltonian reads: 
\begin{equation}
H=H_{s}+\sum_{ij\in\left\{ 12,23,13\right\} }\left(H_{b}^{ij}+f_{s}^{ij}g_{b}^{ij}\right).\label{eq:total_hamil}
\end{equation}
The system part $H_{s}=\sum_{k=1}^{3}E_{k}\left|k><k\right|$ is simple
diagonal with energy levels $E_{k}$. Each bath is a collection of
harmonic oscillators $H_{b}^{ij}=\sum_{\alpha}\left[\frac{1}{2}m_{\alpha}^{ij}\left(\omega_{\alpha}^{ij}x_{\alpha}^{ij}\right)^{2}+\frac{\left(p_{\alpha}^{ij}\right)^{2}}{2m_{\alpha}^{ij}}\right]$,
where $\alpha$ counts the oscillators of a bath and $ij$ refers
to the two system levels exchanged by the said bath. The interaction
rises between the collective coordinate of the bath oscillators $g_{b}^{ij}=\sum_{\alpha}c_{\alpha}^{ij}x_{\alpha}^{ij}$
and the transition of the system $f_{s}^{ij}=\left|i><j\right|+c.c.$

The same reasoning as for Eq. (\ref{eq:energy_flux_j}) leads to the
expression of the energy fluxes 
\begin{equation}
\left\langle J^{ij}\right\rangle =\frac{i}{\hbar}Tr_{s}\left\{ \left[f_{s}^{ij},H_{s}\right]\tilde{\rho}_{s,1}^{ij}\right\} ,\label{eq:flux-3bath-1}
\end{equation}
which directly verifies the energy conservation $\frac{d}{dt}Tr_{s}\left(\tilde{\rho}_{s}H_{s}\right)=\sum_{ij\in\left\{ 12,23,13\right\} }\left\langle J^{ij}\right\rangle $.
$\tilde{\rho}_{s,1}^{ij}$ is defined similar to Eq. (\ref{eq:redu_rho})
as 
\[
\tilde{\rho}_{s,1}^{ij}=M_{u_{1},u_{2}}\left\{ \bar{\rho}_{s}\left(t\right)\bar{g}^{ij}\left(t\right)\right\} ,
\]
where $\bar{g}^{ij}\left(t\right)$ is the random force caused by
the heat bath $H_{b}^{ij}$. $\bar{\rho}_{s}\left(t\right)$ is determined
by a SDE similar to Eq. (\ref{eq:sto_rho}).

The same manipulation briefed in the previous section transforms the
SDE of $\bar{\rho}_{s}$ to HEOM (see Appendix \ref{sec:HEOM-For-A}
and Ref. \cite{yan2012lasercontrol}). We utilize HYSHE package to
solve the coupled differential equations of HEOM. HYSHE is designed
with optimized data structure to solve quantum dynamics of small systems
coupled to multiple Caldeira-Leggett baths and time-dependent driven
forces. Many truncation policies are implemented at ready to accelerate
the convergence. We adopt Debye-cutoff $\Omega$-form spectral density
function 
\begin{equation}
J\left(\omega\right)=\eta\omega\frac{1}{\left[1+\left(\omega/\omega_{c}\right)^{2}\right]}\label{eq:spectral_density}
\end{equation}
readily provided in the package for all three baths. As the original
program only records the first layer of the hierarchy (i.e. the reduced
density matrix), some lines are amended to capture and process the
second layer terms for the deduction of energy fluxes. The modification
is, generally saying, minimal.

\subsection{The parameters of the work dump\label{subsec:The-parameterization-of} }

\begin{figure}
\includegraphics[scale=0.42]{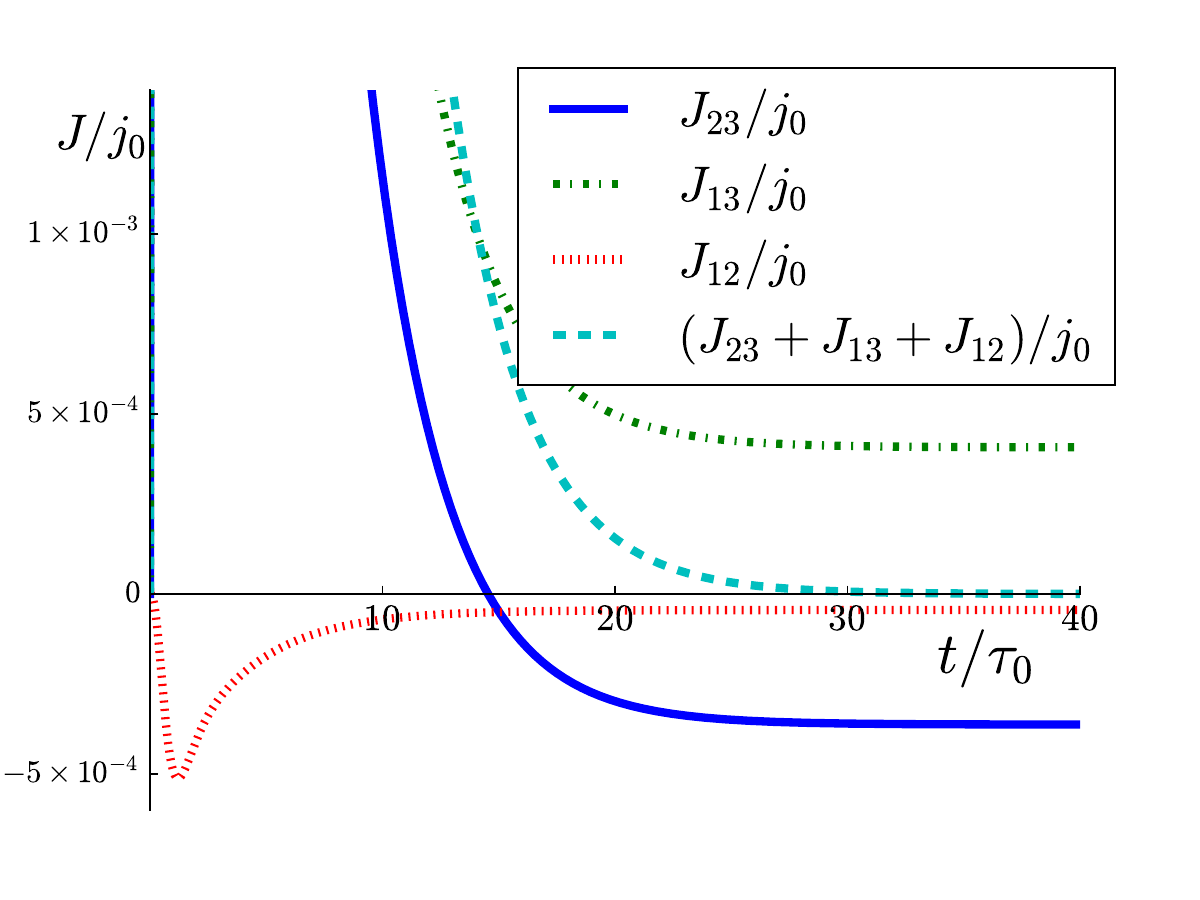}\caption{The fluxes $J/j_{0}$ versus the time $t/\tau_{0}$, the parameters
for the simulation are $E_{1}/\epsilon_{0}=0$, $E_{2}/\epsilon_{0}=0.1$,
$E_{3}/\epsilon_{0}=0.92$, $T_{12}/T_{0}=40$, $T_{23}/T_{0}=3$,
$T_{13}/T_{0}=4$, $\eta_{12}/\eta_{0}=0.025$, $\eta_{13}/\eta_{0}=\eta_{23}/\eta_{0}=0.02$
and $\omega_{c}\tau_{0}=2$ for all three baths.\label{fig:The-fluxes}}
\end{figure}

The immediate outputs of HYSHE are time-dependent bath-specific energy
fluxes. Their typical looks are like Figure \ref{fig:The-fluxes}.
In the plot, the unit of temperature is set to be $T_{0}=75\text{K }$
and so the units of energy, time, energy flux and friction are derived,
respectively, as $\epsilon_{0}=k_{B}T_{0}=1.035\times10^{-21}\text{J}$,
$\tau_{0}=\frac{\hbar}{\epsilon_{0}}=1.019\times10^{-13}\text{s}$,
$j_{0}=\frac{\epsilon_{0}}{\tau_{0}}=1.016\times10^{-8}\text{J/s}$
and $\eta_{0}=\epsilon_{0}\tau_{0}=\hbar$. The unit of the friction
constant $\eta$ is also $\epsilon_{0}$. We choose the solely occupied
ground state of the system to start the propagation. The baths are
initially in their factorized equilibrium as required by HEOM. The
result of a simulation generally displays plateaux of fluxes quickly
reached after a short disturbance caused by initial thermalization.
Once the steady state builds up, the three fluxes cancel out. The
simulation for Figure \ref{fig:The-fluxes} takes less than an hour
to complete on a single core of Xeon E5 CPU. 

Unlike typical engines working in cycles, the three-level engine works
in a steady state. To catch the steady state, the propagation time
is chosen sufficiently long to ensure the effect of the initial state
fades away. The final (steady) values of the fluxes are then recorded
and studied on their relations to the engine parameters.

\begin{figure}
\includegraphics[scale=0.42]{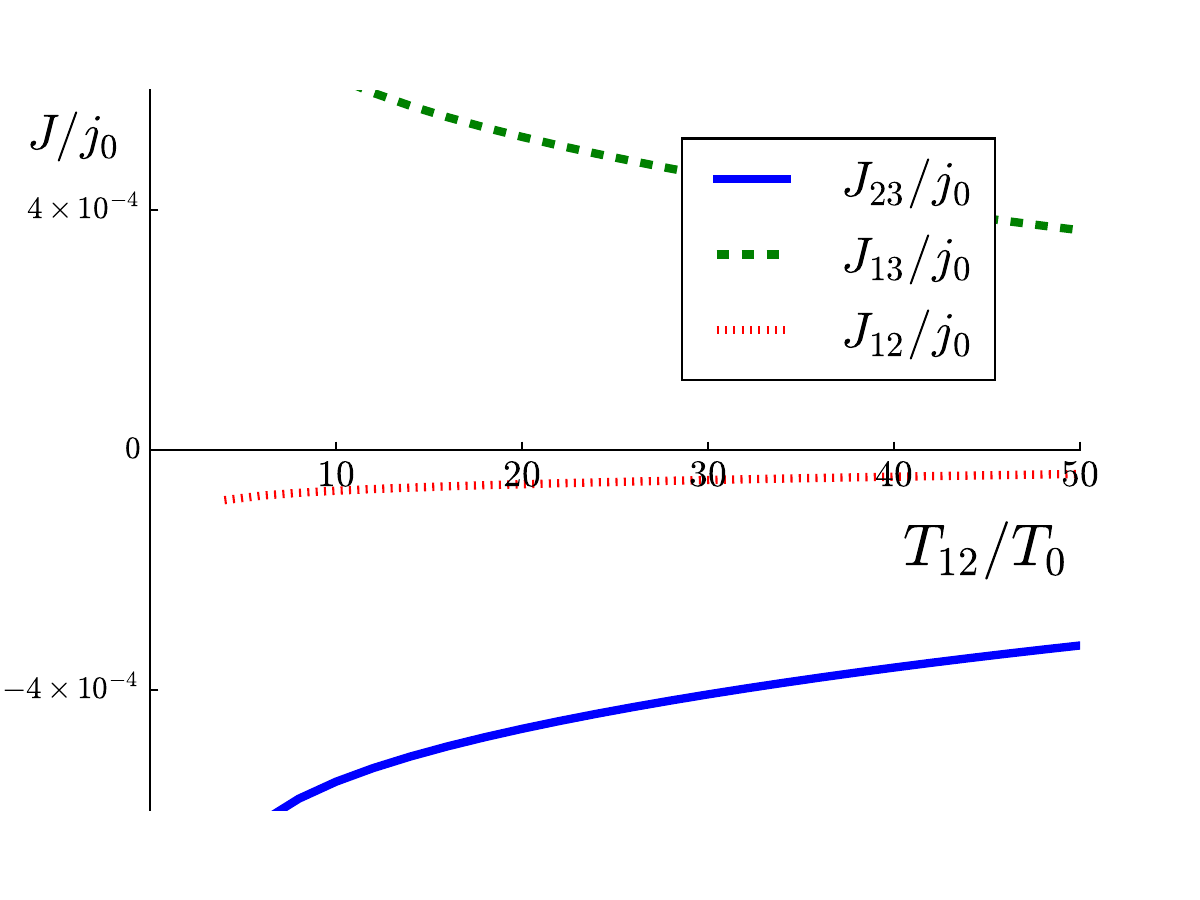}\caption{The fluxes $J/j_{0}$ versus the temperature of the work dump $T_{12}/T_{0}$,
the parameters set for the simulation are $E_{1}/\epsilon_{0}=0$,
$E_{2}/\epsilon_{0}=0.1$, $E_{3}/\epsilon_{0}=0.92$, $T_{23}/T_{0}=3$,
$T_{13}/T_{0}=4$, $\eta_{12}/\eta_{0}=0.025$, $\eta_{13}/\eta_{0}=\eta_{23}/\eta_{0}=0.02$
and $\omega_{c}\tau_{0}=2$ for all three baths. $T_{12}/T_{0}$ is
increased from $4$ to $50$. \label{fig:The-convergence-of}}
\end{figure}

Now that we have three steady fluxes, the next task is to assign different
roles to them by setting up proper parameters to their baths. For
a heat engine, two heat sources with relatively higher and lower temperatures
are mandatory. The fluxes directed to them balance the entropy ($\frac{\delta E_{1}}{T_{1}}+\frac{\delta E_{2}}{T_{2}}=0$)
but leave an energy gap ($\delta E_{1}+\delta E_{2}\neq0$). The gap
should be offset by another work dump bath which takes in proper amount
of energy ($\delta E_{3}=-\left(\delta E_{1}+\delta E_{2}\right)$)
but no entropy. A mechanic way of setting up such a work dump is to
push against a payload. This generally means to introduce a time-dependent
HEOM and some interpretations which we will address in the follow-up
works. In the present work, we choose the thermodynamic way similar
to Ref. \cite{allahverdyan2016adaptive} to set up the work dump.
Simply saying, any energy flowing out of the third bath is accompanied
by entropy $\delta S=\frac{\delta E}{T}$. Here $T$ is the temperature
of the bath, which also implies the assumption of thermal equilibrium
and infinite heat capacity of the bath. It is immediately noticed
that setting $T$ to infinity would meet the ``energy but no entropy''
criteria. In practice the infinitely high temperature can be approximated
by a finite but large enough value relative to those of the other
two baths. However, as is shown in Figure \ref{fig:The-convergence-of},
increasing temperature is not enough by only itself. It depresses
the fluxes without a foreseeable non-zero limit while our work-dump
bath is expected to have a fixed non-zero limit of flux. The reason
is, we hint, that increasing the temperature not only projects the
bath to its limiting work-dump behavior, but also moves the targeted
limit. To see closer, we first refer to the fact that the effect of
the harmonic bath upon our system is completely depicted by the $\Omega$-form
spectral density function $J\left(\omega\right)$ and in turn its
response function $\alpha\left(t\right)=\frac{1}{\pi}\int_{0}^{\infty}d\omega\ \eta\omega C\left(\omega\right)\left[\coth\left(\frac{\hbar\omega}{2k_{B}T}\right)\cos\left(\omega t\right)-i\sin\left(\omega t\right)\right]$.
Here the cutoff function $C\left(\omega\right)$ truncates $J\left(\omega\right)=\eta\omega C\left(\omega\right)$
to finite bandwidth $\omega_{c}$. The limiting work-dump behavior
of the bath requires a finite form of $\alpha\left(t\right)$ when
temperature $T$ goes to infinity. But exclusively increasing temperature
makes an unphysical infinity in the real part of $\alpha\left(t\right)$.
To resolve the singularity we observe that the real part of $\alpha\left(t\right)$
depends on $T$ through $\coth\left(\frac{\hbar\omega}{2k_{B}T}\right)$
which converges to $\frac{2k_{B}T}{\hbar\omega}$ when $T$ is large
enough. This feature prompts that setting the rest engine parameter
$\eta$ proportional to $1/T$ makes the anticipated limit. Without
losing generality, we choose units so that $k_{B}=\hbar=1$ and set
$\eta=\frac{1}{T}$. For $T\gg\omega_{c}$, the real part of $\alpha\left(t\right)$
is reduced to a finite $\alpha_{R}\left(t\right)=\frac{2}{\pi}\int_{0}^{\infty}d\omega\ C\left(\omega\right)\cos\left(\omega t\right)$.
This, along with $\lim_{T\to\infty}\alpha_{I}\left(t\right)=0$, defines
the limiting work-dump behavior of the bath. The finite limit of $\alpha\left(t\right)$
promises a non-zero intake of energy and $T\to\infty$ ensures no
entropy is attached to the energy. This makes the nature of a work
dump. We are aware that the work dump has a characteristic time scale
$\omega_{c}^{-1}$. Though there is still not a ubiquitous definition
of work consistent throughout all scenarios, it is not surprising
that the expected one should be specific upon its time scale. For
example, femtosecond time resolution allows one to track the forces
and displacements of molecules and estimate the work as their products,
while time scale of a second blurs out all details and leaves only
heat flux to be observed. To name the same movement work or heat one
must specify its time scale beforehand.

\begin{figure}
\includegraphics[scale=0.42]{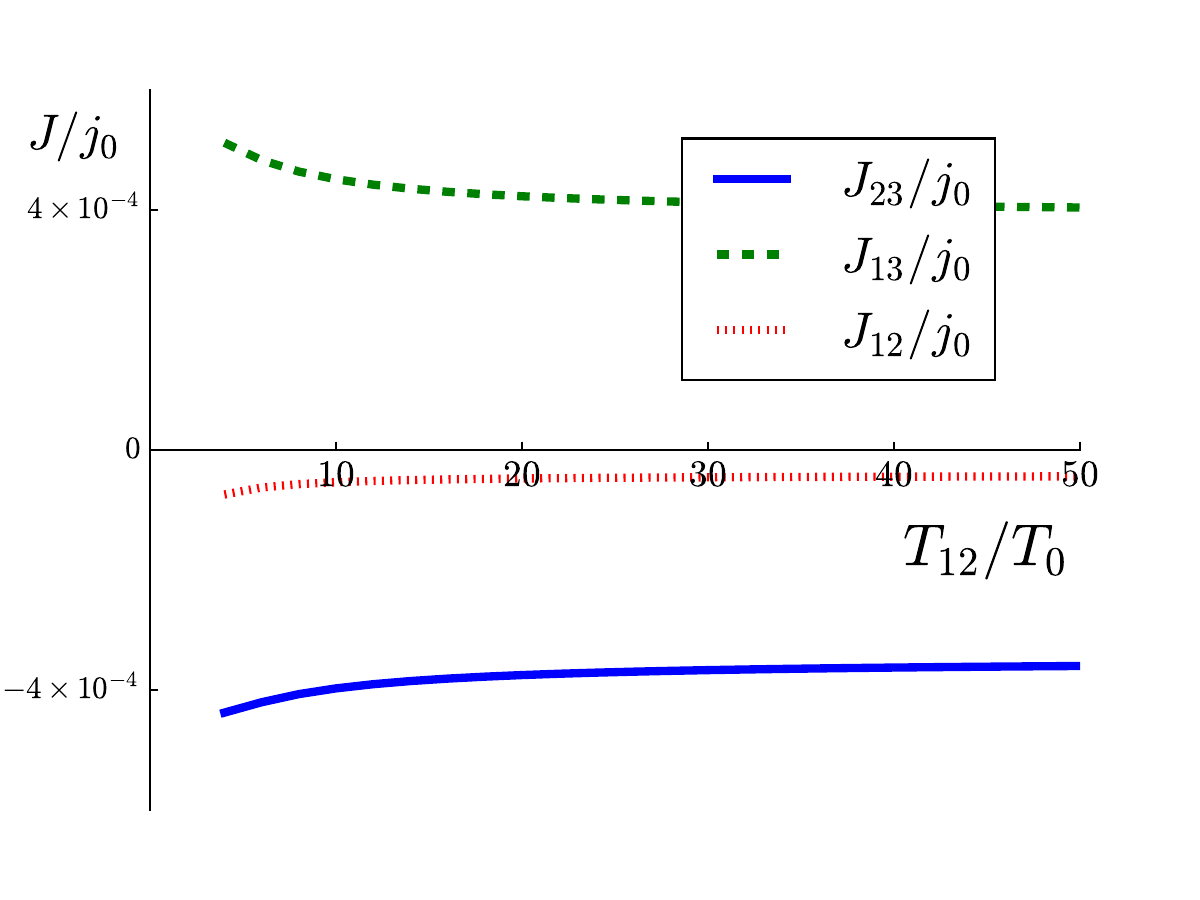}

\caption{The fluxes $J/j_{0}$ versus the temperature of the work dump $T_{12}/T_{0}$,
the parameters set for the simulation are $E_{1}/\epsilon_{0}=0$,
$E_{2}/\epsilon_{0}=0.1$, $E_{3}/\epsilon_{0}=0.92$, $T_{23}/T_{0}=3$,
$T_{13}/T_{0}=4$, $\eta_{13}/\eta_{0}=\eta_{23}/\eta_{0}=0.02$ and
$\omega_{c}\tau_{0}=2$ for all three baths. $T_{12}/T_{0}$ is increased
from $4$ to $50$ and $\eta_{12}/\eta_{0}=\frac{1}{\left(T_{12}/T_{0}\right)}$.\label{fig:The-fluxes-converged}}
\end{figure}

In the simulation we take the third bath as the work dump, increase
its temperature $T_{12}$ and set $\eta_{12}/\eta_{0}=\frac{1}{T_{12}/T_{0}}$
accordingly while keeping the parameters of the other two baths invariant.
The result is displayed in Figure \ref{fig:The-fluxes-converged}.
The plateaux in the figure suggest the work-dump limit be reached
when $T_{12}$ is much larger than $T_{13}$ and $T_{23}$. In the
following simulations we choose $T_{12}$ large enough to safely converge
the third bath and study the effect of other parameters. 

\section{The efficiency of a three-level heat engine\label{sec:The-efficiency-of}}

\begin{figure}
\includegraphics[scale=0.42]{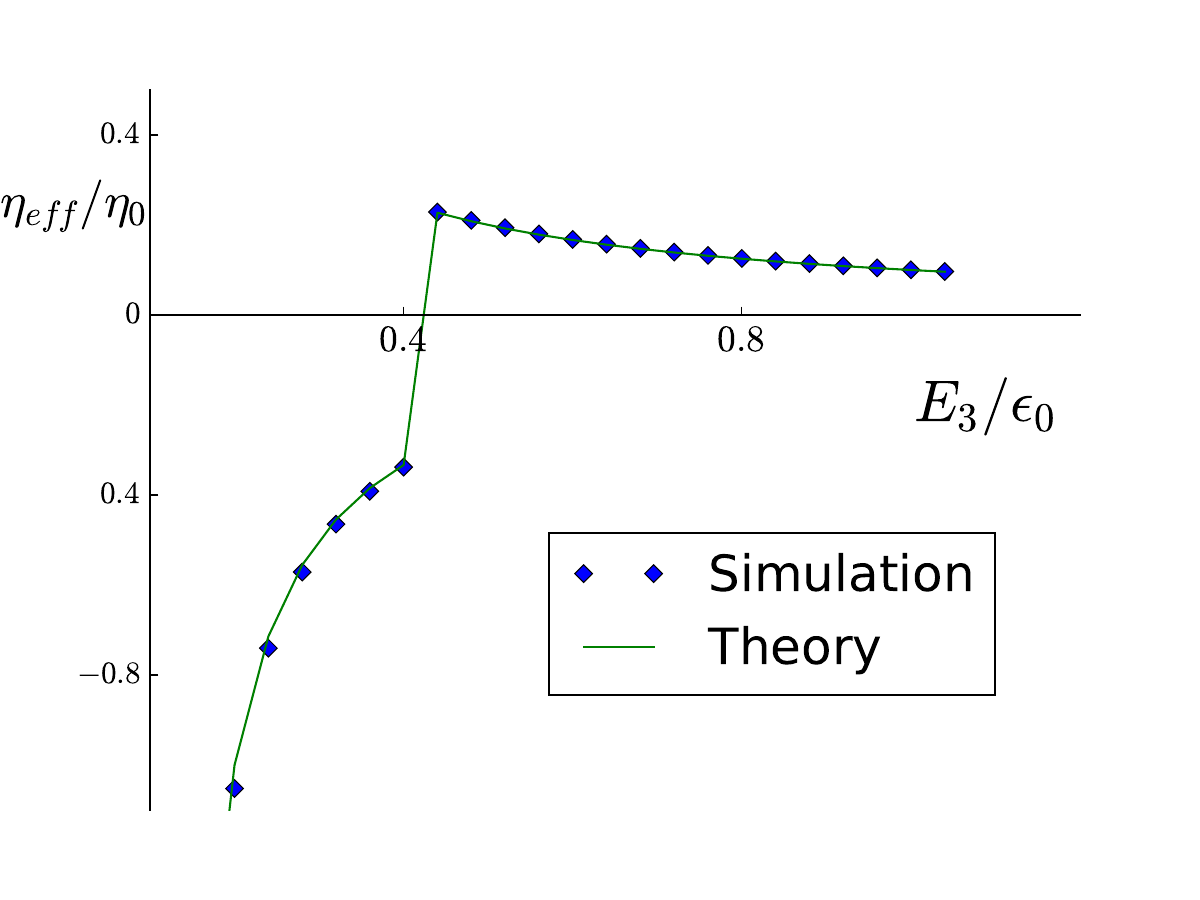}\caption{The efficiency of the heat engine $\eta_{eff}/\eta_{0}$ versus $E_{3}/\epsilon_{0}$,
the parameters set for the simulation are $E_{1}/\epsilon_{0}=0$,
$E_{2}/\epsilon_{0}=0.1$, $T_{12}/T_{0}=40$, $T_{23}/T_{0}=3$,
$T_{13}/T_{0}=4$, $\eta_{12}/\eta_{0}=0.025$, $\eta_{13}/\eta_{0}=\eta_{23}/\eta_{0}=0.02$
and $\omega_{c}\tau_{0}=2$ for all three baths. $E_{3}/\epsilon_{0}$
is increased from $0.2$ to $1.04$.\label{fig:-versus-the}}
\end{figure}

As an application of the technique proposed above, we study the effect
of the parameters on the efficiency of the engine. To isolate the
effect, the parameters related to the work dump need to be fixed first,
including $T_{12}/T_{0}=40$, $\eta_{12}/\eta_{0}=0.025$, $E_{1}/\epsilon_{0}=0$
and $E_{2}/\epsilon_{0}=0.1$. It leaves us only the parameters of
the other two baths and $E_{3}$ to manipulate. We choose to move
the only ``internal'' parameter $E_{3}/\epsilon_{0}$ from $0.2$
to $1.04$ and observe its effect on the efficiency of the engine.
The other parameters are set as $T_{13}/T_{0}=4$, $T_{23}/T_{0}=3$
and $\eta_{13}/\eta_{0}=\eta_{23}/\eta_{0}=0.02$. The cutoffs of
all three baths are set uniformly as $\omega_{c}\tau_{0}=2.0$. The
steady fluxes $J_{12}$, $J_{23}$ and $J_{13}$ are taken from the
levels of the plateaux and then fed into the definition $\eta_{eff}=\frac{-J_{12}}{Max\left(J_{13},J_{23}\right)}$(see
Ref. \cite{allahverdyan2016adaptive}) for the efficiency of the heat
engine. For the small coupling strength we choose, the shifts of the
energy gaps are negligible, thus the energy flux could be approximated
by $I\left(E_{i}-E_{j}\right)$ where $I$ is the population circulation.
The efficiency $\eta_{eff}=\frac{I\left(E_{2}-E_{1}\right)}{Max\left(I\left(E_{3}-E_{1}\right),I\left(E_{2}-E_{3}\right)\right)}$
is then found to be independent of the magnitude of $I$. The sign
of $I$ is inferred from the sign of the energy flux $J_{ij}$. This
theory is compared to the numerical simulation in Figure \ref{fig:-versus-the}
and the results fit well to each other. Note that the gap in the efficiency
$\eta_{eff}$ is caused by the switch from refrigerator to heat engine.

\section{Conclusion}

In the framework of stochastic decoupling and for the bath model comprised
of harmonic oscillators, we find the expression of energy flux to
be determined by the correlation between the random density matrix
of the system and the random force exerted by the bath (Eq. (\ref{eq:redu_rho})
and (\ref{eq:energy_flux_j})). Through the connection between stochastic
decoupling and HEOM we further express the energy flux as a combination
of the terms in the second layer of HEOM. The result makes an efficient
and numerically exact approach that calculates energy fluxes accurately.
We argue that to convert a bath to a work dump one needs to increase
the temperature $T$ of the bath to a large enough value and decrease
the friction coefficient $\eta$ accordingly to keep $\eta T$ invariant.
With this efficient method we calculate the efficiency of a three-level
heat engine weakly coupled to its heat baths. The result compares
well to the prediction of a simple theory for weak coupling limit. 
\begin{acknowledgments}
The work is supported by National Natural Science Foundation of China(NSFC)
under Grant No. 21503048 and 21373064, Program to Support Excellent
Innovative Scholars from Universities of Guizhou Province in Scientific
Researches under Grant No. QJH-KY{[}2015{]}483, Science and Technology
Foundation of Guizhou Province under Grant No. QKH-JC{[}2016{]}1110.
The computation is carried on the facilities of National Supercomputer
Center in Guangzhou, China. 
\end{acknowledgments}

\appendix

\section{Energy Fluxes For A Three-Level System \label{sec:HEOM-For-A}}

With the help of stochastic decoupling, the SDEs for the system and
the baths are shown as, respectively, 
\[
i\hbar d\rho_{s}\left(t\right)=\left[H_{s},\rho_{s}\left(t\right)\right]dt+\frac{1}{2}\sum_{ij}\left[f_{s}^{ij},\rho_{s}\left(t\right)\right]u_{1,t}^{ij}dt+\frac{i}{2}\sum_{ij}\left\{ f_{s}^{ij},\rho_{s}\left(t\right)\right\} u_{2,t}^{ij*}dt
\]
\[
i\hbar d\rho_{b}^{ij}\left(t\right)=\left[H_{b}^{ij},\rho_{b}^{ij}\left(t\right)\right]dt+\frac{1}{2}\left[g_{b}^{ij},\rho_{b}^{ij}\left(t\right)\right]u_{2,t}^{ij}dt+\frac{i}{2}\left\{ g_{b}^{ij},\rho_{b}^{ij}\left(t\right)\right\} u_{1,t}^{ij*}dt.
\]
Here $ij\in\Omega=\left\{ 12,23,13\right\} $. The total density matrix
is expressed as the average $\rho\left(t\right)=M\left\{ \rho_{s}\prod_{ij\in\Omega}\rho_{b}^{ij}\right\} $.
Here the operator $M\left\{ \cdot\right\} $ averages its operand
over all noises introduced by stochastic decoupling. Applying Itô
calculus to the derivative of the average reproduces Liouville equation
$i\hbar\frac{d}{dt}\rho=\left[H,\rho\right]$, which demonstrates
the validity of the SDEs.

Similar to Eq. (\ref{eq:flux_expr}), the fluxes are defined by the
increment of the system energy $\frac{d}{dt}\left\langle H_{s}\right\rangle =-\frac{i}{\hbar}\sum_{ij}\left\langle g_{b}^{ij}\left[H_{s},f_{s}^{ij}\right]\right\rangle =\sum_{r=1}^{3}\left\langle J^{ij}\right\rangle $.
With the stochastic expression of $\rho$, the flux $\left\langle J^{ij}\right\rangle =\frac{i}{\hbar}Tr\left\{ \rho g_{b}^{ij}\left[f_{s}^{ij},H_{s}\right]\right\} $
is simplified to an expression in the finite-dimensional reduced space
\begin{equation}
\left\langle J^{ij}\right\rangle =\frac{i}{\hbar}Tr_{s}\left\{ \left[f_{s}^{ij},H_{s}\right]\tilde{\rho}_{s,1}^{ij}\right\} \label{eq:flux-3bath}
\end{equation}
where 
\begin{align}
\tilde{\rho}_{s,1}^{ij} & =M\left\{ \rho_{s}\bar{g}^{ij}\left(t\right)\prod_{i^{'}j^{'}\in\Omega}Tr_{b}^{i^{'}j^{'}}\left\{ \rho_{b}^{i^{'}j^{'}}\right\} \right\} \label{eq:red_mat_3_bath}\\
\tilde{\rho}_{s,0} & =M\left\{ \rho_{s}\prod_{i^{'}j^{'}\in\Omega}Tr_{b}^{i^{'}j^{'}}\left\{ \rho_{b}^{i^{'}j^{'}}\right\} \right\} .\label{eq:red_mat_3_bath_1}
\end{align}
Here $\bar{g}^{ij}\left(t\right)=\sqrt{\hbar}\int_{0}^{t}\alpha_{I}^{ij}\left(t-\tau\right)u_{2,t^{'}}^{ij}dt+\sqrt{\hbar}\int_{0}^{t}\alpha_{R}^{ij}\left(t-\tau\right)u_{1,t^{'}}^{ij*}dt$
is the random force caused by the bath $H_{b}^{ij}$. Girsanov transformation
absorbs the trace of the bath density matrices in Eq. (\ref{eq:red_mat_3_bath},\ref{eq:red_mat_3_bath_1})
and reformulates the equations as $\tilde{\rho}_{s,0}=M\left\{ \bar{\rho}_{s}\right\} $
and $\tilde{\rho}_{s,1}^{ij}=M\left\{ \bar{\rho}_{s}\left(t\right)\bar{g}^{ij}\left(t\right)\right\} $.
Here $\bar{\rho}_{s}$ is determined by $i\hbar d\bar{\rho}_{s}=\left[H_{s},\bar{\rho}_{s}\right]dt+\sum_{ij}\left[f_{s}^{ij},\bar{\rho}_{s}\right]\bar{g}^{ij}\left(t\right)dt+\sum_{ij}\left\{ \frac{1}{2}\left[f_{s}^{ij},\bar{\rho}_{s}\right]u_{1}^{ij}dt+\frac{i}{2}\left\{ f_{s}^{ij},\bar{\rho}_{s}\right\} u_{2}^{ij*}dt\right\} $.
Averaging the SDE, we have $i\hbar d\tilde{\rho}_{s}=\left[H_{s},\tilde{\rho}_{s}\right]dt+\sum_{ij}\left[f_{s}^{ij},\rho_{s,1}^{ij}\right]dt$,
which easily verifies 
\begin{equation}
\frac{d}{dt}Tr_{s}\left(\tilde{\rho}_{s}H_{s}\right)=\sum_{r}\left\langle J^{ij}\right\rangle .\label{eq:3_fluxes}
\end{equation}

Though in practice HYSHE package itself takes the burden of the job,
we would like to bring forward the equations playing behind the codes
to add some transparency to the black box. For convenience, we ommit
the bath index $ij$ for now. For the chosen spectral density function
Eq. (\ref{eq:spectral_density}), the response function is split as:
\begin{align}
\alpha\left(t\right) & =\frac{\eta\omega_{c}^{2}}{2}\left[\cot\left(\beta\hbar\omega_{c}/2\right)-i\right]\exp\left(-\omega_{c}t\right)\label{eq:response_function}\\
 & +\frac{2\eta\omega_{c}^{2}}{\hbar\beta}\sum_{k=1}^{\infty}\frac{v_{k}\exp\left(-v_{k}t\right)}{v_{k}^{2}-\omega_{c}^{2}}\nonumber 
\end{align}
here $v_{k}=2\pi k/\left(\hbar\beta\right)$. We denote for clarity
the coefficients $C_{0}=\sqrt{\hbar}\frac{\eta\omega_{c}^{2}}{2}\cot\left(\beta\hbar\omega_{c}/2\right)$,
$D_{0}=-\sqrt{\hbar}\frac{\eta\omega_{c}^{2}}{2}$ and $\omega_{0}=\omega_{c}$,
$C_{J}=\sqrt{\hbar}\frac{2\eta\omega_{c}^{2}}{\hbar\beta}\frac{v_{J}}{v_{J}^{2}-\omega_{c}^{2}}$
and $\omega_{J}=v_{J}$ for $J>0$, thus the response function is
split as $\sqrt{\hbar}\alpha\left(t\right)=iD_{0}\exp\left(-\omega_{0}t\right)+\sum_{J=0}^{N}C_{J}\exp\left(-\omega_{J}t\right)$.
Respectively $\bar{g}\left(t\right)$ is expressed as 
\begin{align}
\bar{g}\left(t\right) & =\sqrt{\hbar}\int_{0}^{t}\alpha_{I}\left(t-\tau\right)u_{2,\tau}d\tau+\sqrt{\hbar}\int_{0}^{t}\alpha_{R}\left(t-\tau\right)u_{1,\tau}^{*}d\tau\label{eq:random_force}\\
 & =\int_{0}^{t}D_{0}\exp\left(-\omega_{0}\left(t-\tau\right)\right)u_{2,\tau}d\tau+\sum_{J=0}^{N}\int_{0}^{t}C_{J}\exp\left(-\omega_{J}\left(t-\tau\right)\right)u_{1,t^{'}}^{*}d\tau\nonumber \\
 & =\sum_{J}\bar{g}_{J}\left(t\right)\nonumber 
\end{align}
where $\bar{g}_{J}\left(t\right)=\delta_{J,0}\int_{0}^{t}D_{0}\exp\left(-\omega_{0}\left(t-\tau\right)\right)u_{2,\tau}d\tau+\int_{0}^{t}C_{J}\exp\left(-\omega_{J}\left(t-\tau\right)\right)u_{1,\tau}^{*}d\tau$.
Its derivative is $d\bar{g}_{J}\left(t\right)=\delta_{J,0}D_{0}u_{2,t}dt+C_{J}u_{1,t}^{*}dt-\left(\delta_{J,0}\omega_{0}\bar{g}_{0,im}\left(t\right)+\omega_{J}\bar{g}_{J,re}\left(t\right)\right)dt$
where $\bar{g}_{0,im}\left(t\right)=\int_{0}^{t}D_{0}\exp\left(-\omega_{0}\left(t-\tau\right)\right)u_{2,\tau}d\tau$,
and $\bar{g}_{J,re}\left(t\right)=\int_{0}^{t}C_{J}\exp\left(-\omega_{J}\left(t-\tau\right)\right)u_{1,\tau}^{*}d\tau$.

To avoid confusion the bath index $ij\in\left\{ 23,13,12\right\} $
will be replaced by $r\in\left\{ 1,2,3\right\} $ hereafter. In this
notation we define $\tilde{\rho}_{I}=M\left\{ \bar{\rho}_{s}\prod_{r=1}^{3}\prod_{J=0}^{N}\left(\bar{g}_{J}^{r}\right)^{I_{r,J}}\right\} $,
where the subscript $I$ itself is a matrix of non-negative integral
entries. Utilizing Itô calculus and averaging over all auxiliary noises
we get 
\begin{align}
i\hbar\frac{d}{dt}\tilde{\rho}_{I} & =-i\sum_{m=1}^{3}\sum_{n=0}^{N}I_{m,n}\omega_{n}^{m}\tilde{\rho}_{I}+\left[H_{s},\tilde{\rho}_{I}\right]+\sum_{m=1}^{3}\sum_{n=0}^{N}\left[f_{s}^{m},\tilde{\rho}_{I+\Delta_{m,n}}\right]\label{eq:HEOM}\\
 & +i\sum_{m=1}^{3}I_{m,0}D_{0}^{m}\left\{ f_{s}^{m},\tilde{\rho}_{I-\Delta_{m,0}}\right\} dt+\sum_{m=1}^{3}\sum_{n=0}^{N}I_{m,n}C_{n}^{m}\left[f_{s}^{m},\tilde{\rho}_{I-\Delta_{m,n}}\right].\nonumber 
\end{align}
Here $\Delta_{m,n}$ is a matrix defined by its entries $\left(\Delta_{m,n}\right)_{r,J}=\delta_{mr}\delta_{nJ}$.
To calculate the energy flux from bath $r$, any term $\rho_{I}$
with all $I$'s entries zero, except one single $1$ at $I$'s $r$'th
row, will be summed up to 
\begin{equation}
\tilde{\rho}_{s,1}^{r}=\sum_{n}\tilde{\rho}_{\Delta_{r,n}}.\label{eq:hie_1st_layer}
\end{equation}
This expression, along with Eq. (\ref{eq:flux-3bath}), calculates
the energy flux from the $r$'th bath.

In HYSHE proper factors are multiplied to scale the hierarchical terms
for better numerical performance: 
\begin{equation}
P_{I}=\frac{\prod_{r}\prod_{J=0}^{N}\left(\omega_{J}^{r}\right)^{I_{r,J}}}{\left(\prod_{r}\prod_{J=0}^{N}\left(I_{r,J}!\right)\left|C_{J}^{r}+i\delta_{J0}D_{J}^{r}\right|^{I_{r,J}}\right)^{\frac{1}{2}}}\tilde{\rho}_{I}.\label{eq:rescale}
\end{equation}
The scaling also changes the hierarchical equations to 
\begin{align}
i\hbar\frac{d}{dt}P_{I} & =-i\sum_{m=1}^{3}\sum_{n=0}^{N}I_{m,n}\omega_{n}^{m}P_{I}+\left[H_{s},P_{I}\right]\label{eq:rescaled_HEOM}\\
 & +\sum_{m=1}^{3}\sum_{n=0}^{N}\left(\omega_{n}^{m}\right)^{-1}\left(I_{m,n}+1\right)^{\frac{1}{2}}\left|C_{n}^{m}+i\delta_{n0}D_{0}^{m}\right|^{\frac{1}{2}}\left[f_{s}^{m},P_{I+\Delta_{n,j}}\right]\nonumber \\
 & +\sum_{m=1}^{3}\sum_{n=1}^{N}\omega_{n}^{m}\left(I_{m,n}\right)^{\frac{1}{2}}\left|C_{n}^{m}\right|^{-\frac{1}{2}}C_{n}^{m}\left[f_{s}^{m},P_{I-\Delta_{m,n}}\right]\nonumber \\
 & +\omega_{0}^{m}\left|C_{0}^{m}+iD_{0}^{m}\right|^{-\frac{1}{2}}I_{m,0}^{\frac{1}{2}}\sum_{m=1}^{3}\left\{ iD_{0}^{m}\left\{ f_{s}^{m},P_{I-\Delta_{m,0}}\right\} +C_{0}^{m}\left[f_{s}^{m},P_{I-\Delta_{m,0}}\right]\right\} .\nonumber 
\end{align}
Thus to compensate, its raw output of the second layer terms ought
to be scaled back to $\tilde{\rho}_{I}=P_{I}\frac{\left(\left|C_{j}^{m}+i\delta_{j0}D_{j}^{m}\right|\right)^{\frac{1}{2}}}{\omega_{J}^{m}}$
before Eq. (\ref{eq:hie_1st_layer}) is employed. 

 \bibliographystyle{unsrt}
\bibliography{ref}

\end{document}